\documentclass[10point,reprint,twocolumn,aps,prb]{revtex4-1}
\usepackage{graphicx}
\usepackage{mathtools}
\usepackage{float}
\usepackage{amsmath}
\usepackage{bm}
\usepackage{amssymb}
\usepackage{amsfonts}
\usepackage{dcolumn}
\usepackage{epsfig}
\usepackage{subfigure}
\usepackage{bm}
\begin{document}

\title{Signature of enhanced spin-orbit interaction in the magnetoresistance of LaTiO$_3$/SrTiO$_3$ interfaces on $\delta$-doping}

\author{Shubhankar Das, Z. Hossain and R. C. Budhani}
\email{rcb@iitk.ac.in}
\affiliation{Condensed Matter - Low Dimensional Systems Laboratory, Department of Physics, Indian Institute of Technology, Kanpur - 208016, India}

\date{\today}
\begin{abstract}
We present a study of modulation of spin-orbit interaction (SOI) at the interface of LaTiO$_3$/SrTiO$_3$ by $\delta$-doping with an iso-structural ferromagnetic perovskite LaCoO$_3$. The sheet carrier density at the interface decreases exponentially with $\delta$-doping thickness. We have explored that the spin-orbit scattering time ($\tau_{so}$) can be decreased by nearly 3 orders of magnitude, whereas the inelastic scattering time ($\tau_{i}$) remains almost constant with $\delta$-doping thickness. We have also observed that the $\tau_{i}$ varies almost inversely proportional to temperature and $\tau_{so}$ remains insensitive to temperature, which suggest that the spin relaxation in these interfaces follows D'yakonov-Perel mechanism. The observed in-plane anisotropic magnetoresistance is attributed to the mixing of the spin up and spin down states of d-band at Fermi level due to SOI.
\end{abstract}

\maketitle


The effects of Rashba type spin-orbit interaction (SOI) on the two dimensional electron gas (2DEG) formed at the interface of III-V compound semiconductors and perovskite oxides like LaAlO$_3$ (LAO) and SrTiO$_3$ (STO) are being addressed extensively in recent years.\cite{Dresselhaus,Bychkov,D'yakonov,Elliott,Yafet,Bergmann,Poole,Sacks,Grbic,Nitta,Caviglia,Shalom,Gyanendra,
Shubhankar1,Pramod} One important influence of SOI is on the diffusive transport of charge carriers in a disordered 2D conductor at low temperatures, which otherwise is governed by quantum correction to conductivity derived from weak localization (WL) and electron-electron interaction (EEI). The WL arises from the constructive interference between two time-reversed partial waves of charge carriers which are scattered by the same defects or impurities but travel in opposite direction along the same close trajectory. This leads to the higher probability of carrier backscattering and hence enhancement of the longitudinal resistivity. A perpendicular magnetic field breaks this quantum interference by introducing a phase shift in the counter rotating partial waves and a negative magnetoresistance ensues.\cite{Knap} The SOI, in particular, has a strong influence on WL as it also breaks the quantum interference. This phenomena is known as Weak Antilocalization (WAL), which manifests itself as a positive magnetoresistance at low fields around H = 0.\cite{Grbic} The strength of these two mechanisms are reflected in the inelastic scattering time $\tau$$_{i}$ and spin-orbit scattering time $\tau$$_{so}$, both of which break quantum interference between electronic partial waves. If $\tau$$_{i}$ $<$ $\tau$$_{so}$ and $>$ $\tau$ ($\tau$ is the elastic scattering time), weak localization effect dominates and a negative MR results. However if $\tau$$_{so}$ $<$ $\tau$$_{i}$, a positive MR is predicted at low field which turn into negative MR at a critical field where the coherent quantum interference is at maximum. In a strong SOI regime, where $\tau$$_{so}$ $\ll$ $\tau$$_{i}$ a positive MR is expected over a large range of field.\cite{Poole}

Weak antilocalization was experimentally observed first by Bergmann in thin films of Mg covered with Au which provides the SOI.\cite{Bergmann} The WAL effects have also been seen in semiconductor heterostructures such as inversion layer of indium phosphide and n-type GaAs.\cite{Poole,Sacks} In p-type GaAs/Al$_x$Ga$_{1-x}$As heterostructure exceptionally strong S-O interaction is observed due to the high effective mass of holes.\cite{Grbic} Importantly, the spin-orbit coupling parameter and the spin-splitting energy can be modulated by applying gate voltage in inverted InGaAs/InAlAs heterostructures.\cite{Nitta} Luo et al. have shown that in InAs based heterostructures, the source of zero-field spin-splitting is dominated by the inversion asymmetry at the interface over the bulk crystal structure.\cite{Luo}

The diffusive two-dimensional metal formed at the interface of LaAlO$_3$/SrTiO$_3$ (LAO/STO)\cite{Ohtomo2,Herranz,Salluzzo,Popovic}, LaTiO$_3$/SrTiO$_3$ (LTO/STO)\cite{Ohtomo,Shibuya,Wong,Rastogi} and even ZnO-MgZnO\cite{Makino,Zhang} provides a new playing field to study and modulate SOI by electrostatic gating and interfacial doping. Caviglia et al. have tuned the SOI at LAO/STO interface by electrostatic gating in a backgated configuration over the field range -6000 V/cm to 2000 V/cm at 2 K.\cite{Caviglia} This allowed $\tau_{so}$ to change by 3 orders of magnitude. Recently, Stornaiuolo et al. have shown that the SOI can also be modulated in a sidegated configuration by increasing the gate field to 3000 V/cm and in the temperature range of 0.3 K to 10 K.\cite{Stornaiuolo} Shalom et al. showed that the S-O coupling energy at the LAO/STO interface can be enhanced by applying backgate voltage.\cite{Shalom} Experiments on (001) and (110) oriented STO of LAO/STO interface highlight the role of orbital occupancy on SOI.\cite{Gyanendra} In our previous study, we have revealed that the SOI in the LTO/STO heterostructure can be enhanced significantly by delta ($\delta$)-doping with an iso-structural antiferromagnetic perovskite LaCrO$_3$ at the interface.\cite{Shubhankar1} Similarly, an enhancement in SOI for the case of LaCr$_x$Al$_{1-x}$O$_3$/SrTiO$_3$ interface has been seen on substitutional doping of Chromium at the Al-sites.\cite{Pramod} In order to establish the role of Cr, we have studied magnetotransport in LTO/STO interfaces $\delta$-doped with ferromagnetic perovskite LaCoO$_3$ (LCO). LCO in the bulk form shows transition between different spin state of cobalt. Below 100 K, the Co$^{3+}$ ions in this system are in a low spin state (S = 0) with t$_{2g}^6$ configuration. Above 100 K, the Co$^{3+}$ ions undergo a spin transition to a higher spin state. However, whether the Co$^{3+}$ ions in the high spin state stay at intermediate spin state (S = 1) with t$_{2g}^5$e$_g^1$ configuration\cite{Korotin, Zobel}, high spin state (S = 2) with t$_{2g}^4$e$_g^2$ configuration\cite{Raccah, Haverkort}, or a complex mixture of these two configurations is still under debate. Recent studies on thin film of LCO grown under tensile strain and in 4$\times$10$^{-1}$ mbar oxygen pressure have revealed ferromagnetic transition below $\approx$ 85 K\cite{Fuchs1, Fuchs2}, in marked contrast with the bulk material which does not show long range magnetic order. This ferromagnetic ordering has been attributed to John-Teller distortions induced by the epitaxial strain. The oxygen content in LCO films also controls ferromagnetic ordering. The films deposited in low oxygen pressure ($\leq$ 1$\times$10$^{-2}$ mbar) creates oxygen vacancies, which leaves one extra electron that may transfer to the Co$^{3+}$ ions, forming larger Co$^{2+}$ ions and in turn expand the lattice. As a consequence the tetragonal lattice distortion is reduced which leads to the suppression of ferromagnetic ordering.\cite{Mehta,Biskup} We anticipate that the three extra 3d electrons in Co$^{3+}$ ion and ferromagnetic ordering in LCO will further strengthen the SOI as compared to the SOI in LTO/STO doped with the antiferromagnetic chromate. \cite{Shubhankar1}

The LTO/LCO/STO heterostructures were deposited in a layer-by-layer manner using pulse laser ablation as described as our earlier work.\cite{Shubhankar2} The STO (001) substrate was etched with HF buffered solution to get a TiO$_2$ terminated surface and then annealed at 800$^\circ$C in the growth chamber maintained at 7.4$\times$10$^{-2}$ mbar oxygen for an hour to realize a terraced defects-free surface. The films of LCO and LTO were deposited in 1$\times$10$^{-4}$ mbar oxygen at 800$^\circ$C with the laser fluence and repetition rate of 1.2 J/cm$^2$ and 1 Hz respectively. This resulted in a growth rate of $\sim$ 0.1 \AA/s. The structural analysis of LCO and LTO films are performed by X-ray diffraction (see supplementary material for details).\cite{Supplementary}  Electrical transport measurements were performed in the four probe and Van-der Pauw geometries down to 2 K in a Physical Properties Measurement System (PPMS) equipped with a 14 Tesla ($\mathcal{T}$) superconducting magnet and a precision sample rotator.

\begin{figure}[h!]
\begin{center}
\includegraphics [width=8.5cm]{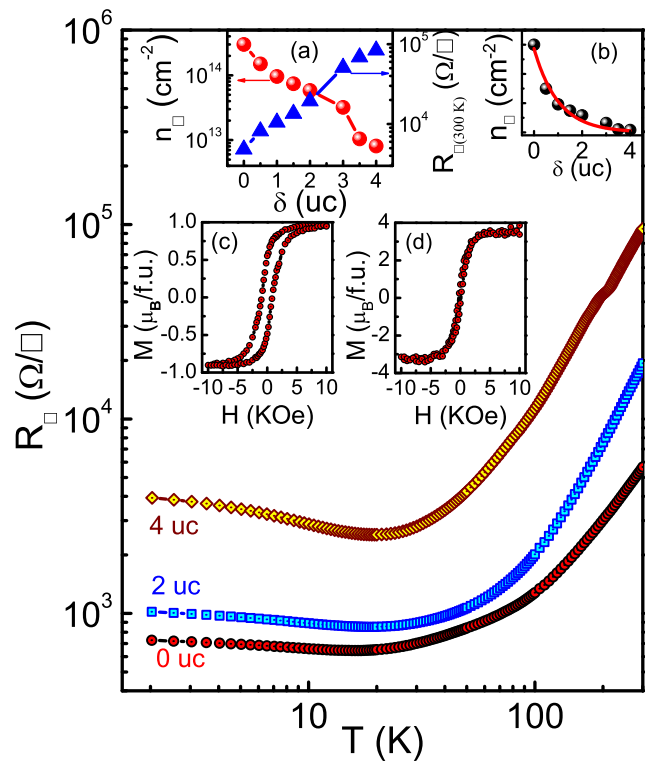}%
\end{center}
\caption{\label{fig1} Temperature dependence sheet resistance of LTO(20uc)/LCO($\delta$-uc)/STO heterostructures; where $\delta$ = 0, 2, 4 uc. Inset (a): the room temperature sheet carrier density and sheet resistance as a function of doping thickness for the same heterostructures. In inset (b) n$_\Box$ is plotted in linear scale and the solid line is fitting using n$_\square  (\delta )$ = n$_\square  (0)e^{ - A\delta }$. Inset (c) and (d): the magnetization hysteresis loop at 5 K of LCO(30 uc)/STO and LTO(20 uc)/LCO(5 uc)/STO heterostructures respectively after subtracting the diamagnetic contribution of the substrate.}
\end{figure}

The strong influence of the $\delta$-layer thickness on electrical transport characteristics of the LTO(20 uc)/LCO($\delta$-uc)/STO heterostructures are displayed in Fig. 1 and its inset for the $\delta$-layer of thickness 0, 2, 4 unit cells (uc). For the undoped LTO/STO sample, the room temperature sheet carrier density (n$_\Box$) is typically 3$\times$10$^{14}$ /cm$^2$.  But it decreases with the $\delta$-layer thickness approximately as n$_\square  (\delta )$ = n$_\square  (0)e^{ - A\delta }$, where A is decay constant with value 1.04. In inset (b) of Fig. 1, the fitting using this equation has been shown. Here it needs to be mentioned that unlike the case of LAO/STO, we have found a linear magnetic field dependence of Hall voltage up to 10 $\mathcal{T}$ for all the doped sample at low temperature. The non-linearity of low temperature Hall voltage in magnetic field has led to the speculation of a hidden magnetic order to explain the low field data and a scenario of multiband conduction to understand the nonlinearity at high fields. Indeed, at LTO/STO interface the nonlinearity in Hall voltage is seen particularly on positive electrostatic gating which sweeps the Ti 3d$_{xy, yz, zx}$ orbital derived bands across the Fermi energy.\cite{Biscaras} In our case, all measurements have been performed under zero gate bias and the Hall resistance remains linear in the field up to 10 $\mathcal{T}$ used here, suggesting that multiband effect may be minimal. The sheet resistance (R$_\Box$) at 300 K increases by two order of magnitude as the $\delta$-layer thickness becomes 4 uc. Since the carrier mobility remains nearly constant with values of 8, 15 and 17 cm$^2$-V$^{-1}$-S$^{-1}$ at 300 K and 71, 76 and 78 cm$^2$-V$^{-1}$-S$^{-1}$ at 2 K for $\delta$ = 0, 2 and 4 uc samples respectively, we can conclude that the emergent insulating behavior on $\delta$-doping is primarily due to the loss of charge carriers. In our earlier studies\cite{Shubhankar1} on LaTiO$_3$/LaCrO$_3$/SrTiO$_3$ heterostructures, a similar rise in R$_\Box$ was seen when the $\delta$ became 10 uc thick. It was further established that the chromate layer absorbs some of the electrons donated by the LTO layers to the interface for the formation of 2DEG. The value of $\left. {\frac{{dn_\square  (\delta )}}
{{d\delta }}} \right|_{\delta  = 0}$ for chromate and cobaltite are -1.23 $\times$ 10$^{14}$ and -2.97 $\times$ 10 $^{14}$ respectively, which suggest that the cobaltite has a higher absorption efficiency than the chromate. In the main panel of Fig. 1, we show the temperature dependence of R$_\Box$ over the range 2 K to 300 K for the three representative samples. The R$_\Box$(T) of the undoped LTO/STO is broadly similar to the behavior reported earlier in other samples of the same class.\cite{Rastogi2} It is characterized by a T$^2$ metallic behavior, followed by a shallow minimum around $\thickapprox$ 50 K and saturation of resistance at still lower temperatures. The latter two features of the R$_\Box$(T) become pronounced in the $\delta$-doped samples. The magnetization hysteresis loop as a function of applied magnetic field ($\pm$2000 Oe) at 5 K of LCO(30 uc)/STO and LTO(20 uc)/LCO(5 uc)/STO heterostructures are plotted in the inset (c) and (d) of fig. 1 respectively. The magnetization data have been collected after cooling the samples at 1000 Oe field. The ferromagnetic phase for both of the samples persists up to 30 K as observed from temperature dependent magnetization curve (not shown in the figure). The lower ferromagnetic ordering temperature in our LCO films can be understood in terms of oxygen vacancies in LCO films created during the deposition.

\begin{figure}[h!]
\begin{center}
\includegraphics [width=8.5cm]{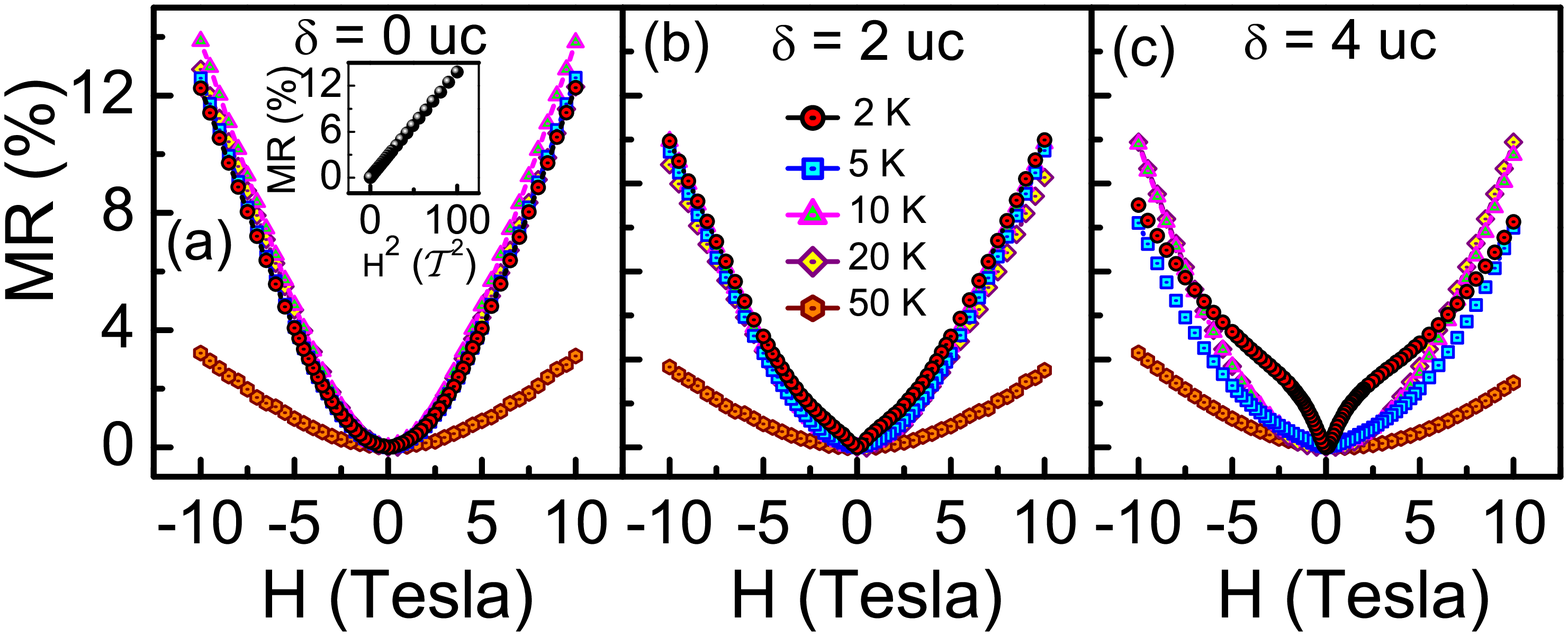}%
\end{center}
\caption{\label{fig2} (a)-(c) The out-of-plane MR of $\delta$ = 0, 2 and 4 uc samples  respectively at various temperatures. While the MR of $\delta$ = 0 uc sample shows quadratic on field at all temperatures, the MR of doped samples deviates from the quadratic behavior at 2 K. Inset of (a): Kohler's plot of $\delta$ = 0 uc sample.}
\end{figure}
We first address the magnetoresistance defined as (MR = $\frac{{R(H) - R(0)}}{{R(0)}}$) of the $\delta$ = 0, 2 and 4 uc samples, as shown in Fig. 2(a)-(c) respectively. These data are for the out-of-plane geometry, where magnetic field is perpendicular to both sample plane and current direction. We label these data as MR$_\perp$. The $\delta$ = 0 uc sample is characterized by a positive MR$_\perp$ at all temperatures and its magnitude increases as we go down to 2 K where it is $\approx$ 14$\%$ at 10 $\mathcal{T}$. This large MR$_\perp$ is attributed to the enhanced transit path and scattering of electrons due to their cyclotron motion in the magnetic field, and it follows the Kohler's rule (MR$_\perp$ $\propto$ aH$^2$) as shown in the inset of Fig. 3(a). The MR$_\perp$ of the $\delta$-doped samples is distinctly lower compared to that of $\delta$ = 0 uc sample for the same value of magnetic field. Moreover, the 2 K MR$_\perp$ of these samples deviates from the quadratic field dependence at lower fields, as indicated by the emergence of a cusp around H = 0 $\mathcal{T}$, which is prominent for $\delta$ = 4 uc sample. The onset of this feature suggests a new scattering phenomena appearing at T $\leqslant$ 5 K. Previous studies have suggested that the cusp-like minimum in MR$_\perp$ is a manifestation of the appearance of strong SOI.\cite{Grbic, Caviglia, Chiu}

\begin{figure}[h!]
\begin{center}
\includegraphics [width=8.5cm]{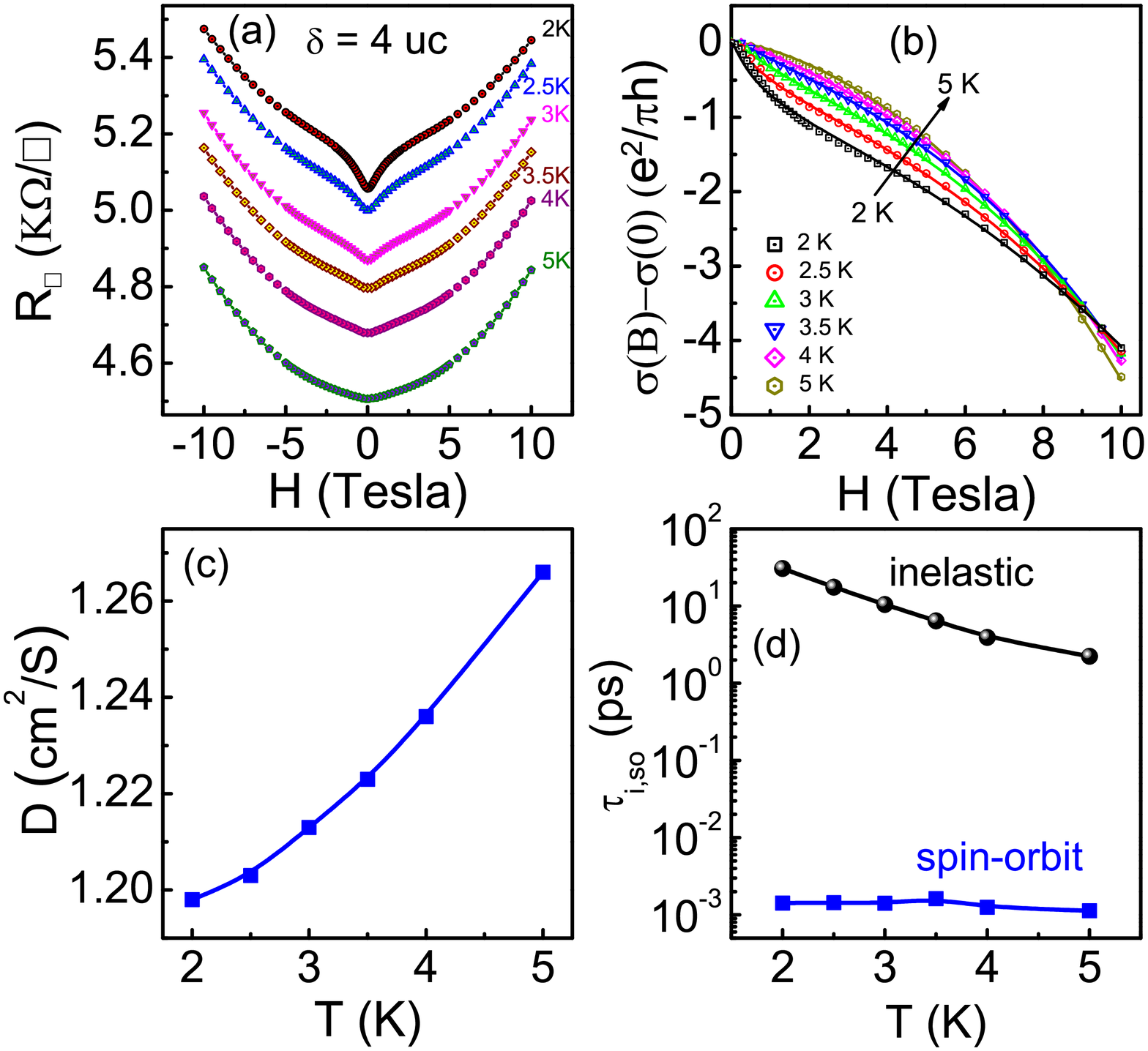}
\end{center}
\caption{\label{fig3} (a) MR curve of $\delta$ = 4 uc sample as a function of out-of-plane magnetic field for various temperatures. (b) The open symbols are the experimental data of conductance correction normalized by the quantum conductance value and the solid lines are the fit using Eq. 1. (c) The diffusion coefficient as a function of temperature for the same sample. (d) Inelastic scattering time $\tau$$_i$ (black circle) and spin-orbit scattering time $\tau$$_{so}$ (blue square) as a function of temperature are plotted on a logarithmic time scale.}
\end{figure}

We now focus on the behavior of this cusp like feature by measuring MR$_\perp$ in smaller temperature intervals below 5 K for the $\delta$ = 4 uc sample. These data are plotted in Fig. 3(a) where we see that the cusp around H = 0 $\mathcal{T}$ diminishes slowly on increasing temperature and it vanishes around T $\sim$ 5 K. The quantum correction to conductivity for a diffusive 2D metal with dominant SOI can be expressed as;\cite{Grbic, Gyanendra, Hikami, Knap, Iordanskii}
\begin{eqnarray*}
\frac{{\Delta \sigma (H)}}
{{G_0 }} &=&  - \left[ {\frac{1}{2}\Psi
\left( {\frac{1}{2} + \frac{{{H_i }}}{H}} \right) - \frac{1}{2}\ln
\frac{{{H_i }}}{H}} \right.\\
&&- \Psi \left( {\frac{1}{2} + \frac{{{H_i } + {H_{so}}}}{H}} \right)
+ \ln \frac{{{H_i } + {H_{so}}}}{H}\\
&&- \frac{1}{2}\Psi \left( {\frac{1}{2} + \frac{{{H_i } +
2{H_{so}}}}{H}} \right) + \ln \left. {\frac{{{H_i } + 2{H_{so}}}}{H}}
\right]\\
&&- A_K \frac{{\sigma (0)}}
{{G_0 }}\frac{{H^2 }}
{{1 + CH^2 }}\;\;\;\;\;\;\;\;\;\;\;\;\;\;\;\;\;\;\;\;\;\;\;\;\;\;\;\;\;\;\;\;\;(1)
\end{eqnarray*}
where $\sigma$ is the longitudinal conductance, obtained from the inversion of experimental resistance data, $\Delta \sigma (H)$ = $\sigma$(H) - $\sigma$(0), $\Psi$(x) is the digamma function, H$_i$ = $\hbar/4eD\tau_i$ and H$_{so}$ = $\hbar/4eD\tau_{so}$ are the characteristics magnetic field, D is the diffusion coefficient and G$_0$ (= e$^2$/$\pi$h) is the quantum of conductance. The last term of Eq. (1) containing A$_K$ and C is the Kohler term which takes into account the classical orbital effect. The open symbols in Fig. 3(b) are the conductance correction at various temperature for $\delta$ = 4 uc sample and the solid lines are the fit using Eq. (1). To extract the relaxation times $\tau_{i}$ and $\tau_{so}$ from H$_i$ and H$_{SO}$, we have calculated the temperature dependence of the diffusion coefficient from the measured sheet carrier density of $\delta$ = 4 uc sample at 2, 2.5, 3, 3.5, 4 and 5 K in van-der pauw geometry. Now an estimation of Fermi velocity (V$_F$ = $\sqrt {2\pi n_\square  } \hbar /m^* $), elastic scattering time ($\tau$ = $m^* \mu /e$) and electron effective mass m$^*$ = 3m$_e$\cite{Caviglia, Mattheiss} where m$_e$ is the mass of the bare electron, the diffusion coefficient can be expressed as D = V$_F$$^2$$\tau$/2.\cite{Caviglia} Fig. 3(c) shows D as a function of temperature for $\delta$ = 4 uc sample. The scattering times $\tau_{i}$ and $\tau_{so}$ are plotted as a function of temperature in Fig. 3(d). While the $\tau_i$ increases nearly by a factor of ten on lowering the temperature from 5 K to 2 K, the $\tau_{so}$ remains constant.

The spin relaxation in a 2D system can occur either through the D'yakonov-Perel' (DP) type process\cite{D'yakonov} or by the Elliot-Yafet (EY) mechanism\cite{Elliott,Yafet}. In the DP process, the SOI arises from the spin splitting of electronic subbands by an electric field whose origin lies in broken inversion symmetry, either in the bulk of the crystal or at the interface. The latter is known as Rashba mechanism. The Rashba SOI hamiltonian is expressed as\cite{Bychkov} $H_{so}  = \alpha \left( {\hat n \times \vec k} \right).\vec S$, where $\alpha$ is S-O coupling constant, $\vec S$ are Pauli matrices, $\vec k$ the Fermi wave vector and $\hat n$ a unit vector perpendicular to the interface. The coupling between the electron spin with the internal magnetic field $\left( {\hat n \times \vec k} \right)$, which is perpendicular to the wave vector and lie in the plane of the interface, is expressed by the Hamiltonian.\cite{Caviglia} In the EY mechanism, the elementary process is the spin-orbit scattering of conduction electrons by the ions of the lattice. In the presence of strong S-O scattering impurity or when the ionic SOI makes significant change to the band structure of the material, this spin relaxation mechanism becomes more dominant. In the DP mechanism, the $\tau_{so}$ remains constant with temperature.\cite{Minkov} On the other hand, in EY mechanism, $\tau_{so}$ decreases on increasing temperature.\cite{Han} In the LTO/LCO/STO heterostructures, the behaviors of $\tau_i$ which is almost inversely proportional to temperature and $\tau_{so}$ is nearly temperature independent, are consistent with the  DP mechanism of spin relaxation.

\begin{figure}[h!]
\begin{center}
\includegraphics [width=8.5cm]{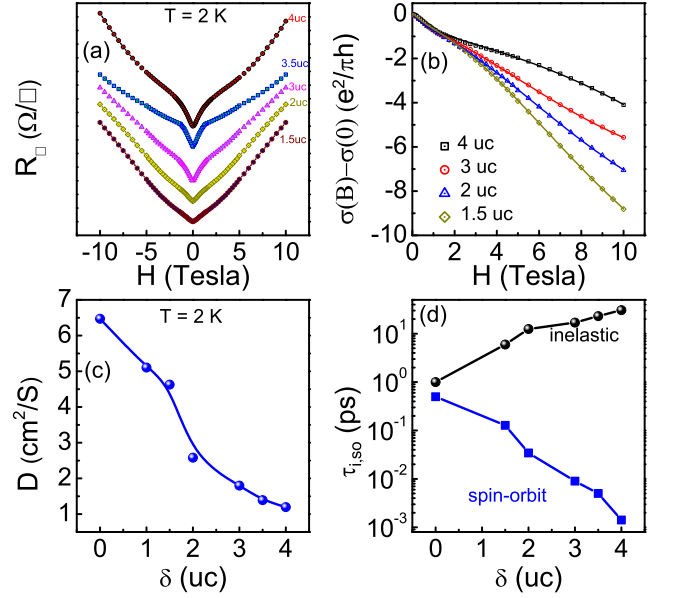}%
\end{center}
\caption{\label{fig4} (a) The MR$_\perp$ of $\delta$ = 1.5, 2, 3, 3.5 and 4 uc sample at 2 K. The scale are being offset for clear vision of the cusp around H = 0 $\mathcal{T}$. (b) Open dots are the conductance peak due to S-O interaction (experimental data) and the solid lines are the fit using Eq. (1). The diffusion coefficient as a function of doping thickness is shown in (c). The modulation of spin-orbit and inelastic scattering time with doping thickness is shown in (d).}
\end{figure}

We have also investigated the $\delta$-layer thickness dependence of the effect of SOI on magnetoresistance. Fig. 4(a) shows the MR$_\perp$ at 2 K of the $\delta$ = 1.5, 2, 3, 3.5 and 4 uc samples. We note that the cusp in MR$_\perp$ at low field emerges clearly with the increasing thickness of the $\delta$-layer, suggesting a much stronger S-O interaction on interface doping. The conductance correction for samples with different $\delta$ and fit to Eq. (1) are shown as open symbols and solid lines respectively in Fig. 4(b). The $\delta$ dependence of diffusion constant and scattering time ($\tau_i$ and $\tau_{so}$) extracted from the fits are shown in Fig. 4(c) and 4(d) respectively. For the $\delta$ = 0 uc sample, where the cusp in MR$_\perp$ is not distinct, the $\tau_i$ and $\tau_{so}$ are nearly the same (0.5-1.0 ps). However, while the former increases marginally on inserting the $\delta$-layer, the $\tau_{so}$ drops by $\approx$ 3 orders of magnitude as $\delta$ reaches 4 uc. A similar change in $\tau_{so}$ at LAO/STO interface has been seen on electrostatic gating in the backgate \cite{Caviglia} and sidegate \cite{Stornaiuolo} configuration. But a remarkable contrast is observed between the effect of electrostatic gating and $\delta$-doping on $\tau_{so}$ if its evolution is compared with the behavior of sheet carrier density. In electrostatic gating, a positive backgate voltage leads to increase of n$_\Box$. Where as in our system, $\delta$-doping at the interface decreases the n$_\Box$. However, the effect of positive gating and $\delta$-doping on SOI and $\tau_{so}$ are similar. This discrepancy can be explained by the multiple-band filling control in SrTiO$_3$ based interfacial 2DEG. In STO-based 2DEG, like LAO/STO and LaVO$_3$/SrTiO$_3$ (LVO/STO), electronic conduction mainly occurs in the Ti 3d derived d$_{xy}$ subbands. On applying a positive backgate voltage electrons start filling the d$_{xz/yz}$ subbands and Fermi energy crosses both the d$_{xy}$ and d$_{xz/yz}$ subbands. It has been shown that in these heterostructures the d$_{xz/yz}$ subbands mainly contribute to the SOI. But the increase of the SOI with d$_{xz/yz}$ subbands filling is not monotonic. The first principle calculations and tight binding analysis indicate that the SOI is largely enhanced at the d$_{xy}$-d$_{xz/yz}$ crossing region due to the orbital mixing.\cite{King,Zhong,Kim,Khalsa,Joshua} In support of these theoretical studies, Liang et al. showed experimentally that in LAO/STO and LVO/STO interfaces the strength of the SOI first increases on increasing the n$_\Box$ by positive gate voltage, followed by a maximum (where n$_\Box$ $\sim$ 3$\times$10$^{13}$ /cm$^2$ at 40 V gate voltage for LAO/STO and n$_\Box$ $\sim$ 4$\times$10$^{13}$ /cm$^2$ at 40 V gate voltage for LVO/STO) and then decrease on further increasing the n$_\Box$.\cite{Liang} As the charge density of pure LTO/STO interface is 1 order of magnitude higher than in the LAO/STO interface used in the above mentioned studies, we expect the Fermi energy for LTO/STO crosses both d$_{xy}$ and d$_{xz/yz}$ subbands. But unlike the case of LAO/STO, the ground state of the LTO/STO interface stays on the opposite side of the d$_{xy}$-d$_{xz/yz}$ crossing region. On decreasing the n$_\Box$ by $\delta$-doping, the system approaches towards the d$_{xy}$-d$_{xz/yz}$ crossing region, and this presumably leads to the enhanced SOI.

\begin{figure}[h!]
\begin{center}
\includegraphics [width=8.5cm]{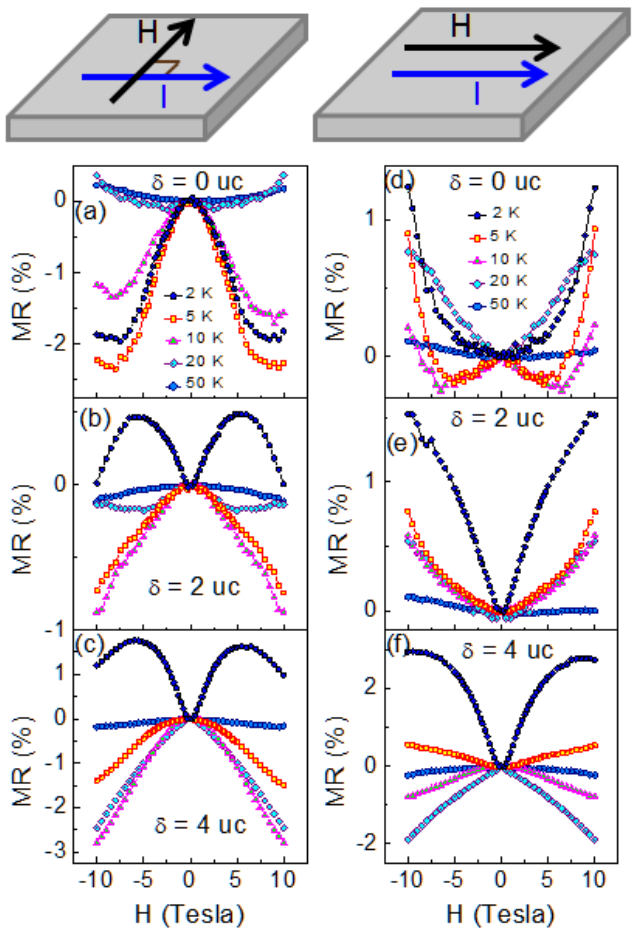}%
\end{center}
\caption{\label{fig5} The in-plane MR of $\delta$ = 0, 2 and 4 uc samples at various temperatures in two different geometries are shown. In first column magnetic field is parallel to the sample plane but perpendicular to the current direction and in last column magnetic field is parallel to both the sample plane and current direction. On the top of two columns the schematic of the geometries are shown.}
\end{figure}

We now present the result of MR measurements performed in a geometry where the magnetic field was in the plane of the 2DEG. In a parallel field, the orbital contribution to magnetoresistance of a thin film becomes negligible, and in fact, vanishes in 2D systems. However, the electrons can still interact with magnetic field via their spin. The MR of $\delta$ = 0, 2 and 4 uc samples in the geometry where the field is in the plane of the sample but aligned perpendicular to the direction of current (MR$_{\parallel\perp}$) are shown in Fig. 5(a)-(c) respectively. Interestingly, a negative MR$_{\parallel\perp}$ for $\delta$ = 0 uc sample is seen for T $<$ 20 K as against the positive MR$_\perp$ (see Fig. \ref{fig2}). This again is suggestive of a new scattering mechanism operational at lower temperatures. Based on our earlier measurements on the LaTiO$_3$/LaCrO$_3$/SrTiO$_3$ heterostructure, the negative MR can be attributed to Kondo effect arising from the interaction between conduction electrons and localized magnetic impurity spins.\cite{Shubhankar1, Shubhankar2} The field and temperature dependence of MR$_{\parallel\perp}$ changes significantly in $\delta$ = 2 and 4 uc samples. At 2 K a positive MR$_{\parallel\perp}$ is seen at low fields which then drops at higher H$_{\parallel\perp}$ resulting in a local maximum at $\sim$ 6 $\mathcal{T}$ for both the samples. One possible source of this positive MR$_{\parallel\perp}$ is the Zeeman interaction of H$_\parallel$ with the conduction electron spin. The major effect of Zeeman interaction is to add a temperature independent dephasing time ($\tau_H$) in the system, which contributes positively to the quantum correction to conductivity due to weak localization. Hence Zeeman interaction destroy antilocalization behavior.\cite{Mal'shukov1} However, this phenomena can be observed only when $\tau_i$ $\gg$ $\tau_H$ or the magnetic field should have been large enough such that $(g\mu _B H)/h \geqslant \left( {\tau _i \tau _{so} } \right)^{ - 1/2}$. But this relation is no longer valid for higher magnetic fields when $(g\mu _B H)/h \geqslant \tau _{so} ^{ - 1}$.\cite{Mal'shukov1,Mal'shukov2} The role of SOI in enhancing Zeeman effect becomes apparent from the absence of positive MR$_{\parallel\perp}$ of $\delta$ = 0 uc sample which has insignificant SOI as indicated by the MR$_\perp$ data of Fig. 2(a).

In order to address the in-plane anisotropy of MR, we have also measured the in-plane MR in a geometry where the external field is parallel to the direction of current. Results of these measurements for $\delta$ = 0, 2 and 4 uc samples are shown in Fig. 5(d)-(f). In this geometry, the in-plane MR (MR$_{\parallel\parallel}$) for $\delta$ = 0 uc at 5 K and 10 K shows negative values at low field but then switch to positive values with increasing field. At 2 K however, the MR$_{\parallel\parallel}$ is positive over the entire field range and it is also higher than MR$_{\parallel\perp}$. Clearly, a large in-plane anisotropy is seen in the MR of these samples. In III-V semiconductor quantum wells, the S-O interaction arises from two contributions; Dresselhaus term\cite{Dresselhaus} and the Rashba term\cite{Bychkov}. In the limiting case, when one of the contributions, either Dresselhaus or Rashba, dominates, the in-plane MR would be isotropic. In other case, when the two contributions are of the same order the in-plane MR would be anisotropic. The degree of anisotropy of the in-plane MR varies with the relative strength of Dresselhaus and Rashba terms.\cite{Mal'shukov1,Mal'shukov2} But in oxide heterostructures, the bulk crystal retains inversion symmetry and hence the Dresselhaus term may not contribute at all to the MR. On the other hand, the ubiquitous anisotropic MR was seen in 3d ferromagnetic transition metals.  To explain this phenomenon, J. Smit\cite{Smit} had proposed that in the presence of SOI, some up spin d-states are mixed into the down spin d-states at the Fermi level. This process allow the s-d scattering to dominate. This mixing of spin up and down d-states are not isotropic because magnetization direction provides a privilege axis for S-O perturbation which leads to the observation of anisotropic magnetoresistance.\cite{Jaoul, Malozemoff} The Smit approach has been used to explain anisotropic MR in LAO/STO interfaces.\cite{Shalom2}

\begin{figure}[h!]
\begin{center}
\includegraphics [width=8.5cm]{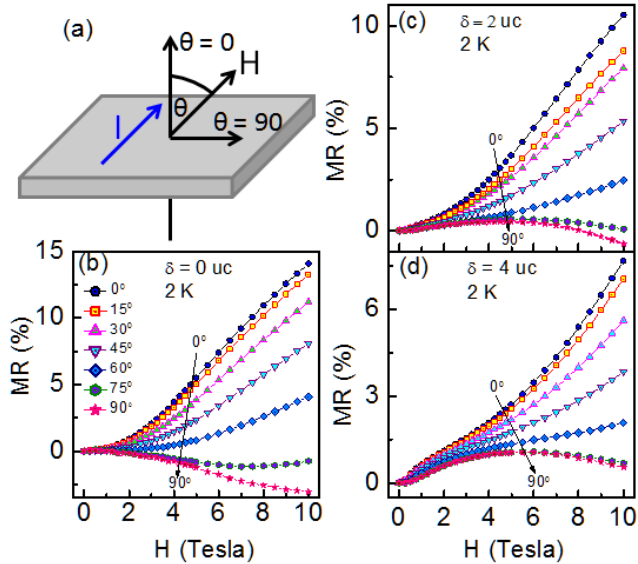}%
\end{center}
\caption{\label{fig6} (a) The schematic diagram which defines the angle ($\theta$) between magnetic field and sample normal. (b)-(d) The MR of $\delta$ = 0, 2 and 4 uc samples respectively at various $\theta$ at 2 K.}
\end{figure}

One may suspect that the positive in-plane MR comes from a slight misalignment (1-2$^{\circ}$) of magnetic field towards out-of-plane direction, which then would give rise to an orbital MR.\cite{Minkov,Mal'shukov2} To investigate this possibility, we have measured the MR of $\delta$ = 0, 2 and 4 uc samples at various angle ($\theta$) between magnetic field and sample plane. Results of such measurements are shown in Fig. 6(b)-(d) respectively, whereas Fig. 6(a) defines the angle $\theta$. The MR$_{\parallel\parallel}$ of $\delta$ = 0 uc sample at 2 K and 10 $\mathcal{T}$ is positive ($\sim$ 1$\%$) (see Fig. 5(d)). A careful look at Fig. 6(b) shows that this amount of positive orbital MR will be generated when $\theta$ becomes $<$ 75$^{\circ}$. Such a gross large misalignment of field is not possible when a precision rotator has been used for the measurement.

\begin{figure}[h!]
\begin{center}
\includegraphics [width=8.5cm]{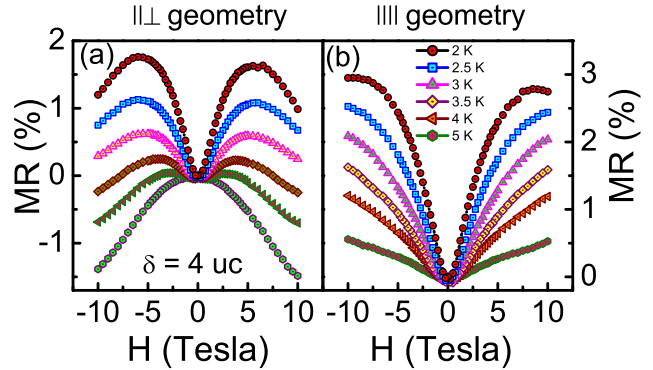}%
\end{center}
\caption{\label{fig7} (a) and (b) The in-plane MR of $\delta$ = 4 uc sample at temperature T = 2, 2.5, 3, 3.5, 4 and 5 K in $\parallel$$\perp$ and $\parallel$$\parallel$ geometry respectively.}
\end{figure}

We also present the MR$_\parallel$$_\perp$ and MR$_\parallel$$_\parallel$ of $\delta$ = 4 uc sample at T = 2, 2.5, 3, 3.5, 4 and 5 K in Fig. 7(a) and (b) respectively. In $\parallel$$\perp$ geometry, the positive MR$_\parallel$$_\perp$ decreases with increasing temperature and at 5 K a negative MR$_\parallel$$_\perp$ is observed. But MR$_\parallel$$_\parallel$ shows a positive value over the entire temperature range 2 K $\leqslant$ T $\leqslant$ 5 K. So the anisotropy of the in-plane MR remains in the whole temperature range 2 K $\leqslant$ T $\leqslant$ 5 K.

In summary, we are successfully able to control the n$_\Box$ of 2DEG at the LTO/STO interface by $\delta$-doping with an iso-structural ferromagnetic perovskite LaCoO$_3$. Here, the Co$^{3+}$ ions at the interface act as traps and absorb electrons which are transferred from LTO to STO side to suppress polar catastrophe. We are also able to enhance the SOI at the LTO/STO interface by the $\delta$-doping technique. A remarkable change ( almost 3 orders of magnitude) in $\tau_{so}$ has been observed by inserting $\delta$ = 4 uc LCO layer at the interface, whereas the change in $\tau_i$ is marginal. We have revealed that in these heterostructures the $\tau_i$ varies nearly as T$^{-1}$, whereas $\tau_{so}$ remains constant with temperature which indicates the spin relaxation follows DP mechanism. The positive in-plane MR has been explained by the Zeeman interaction of external magnetic field with the conduction electron spin. The in-plane anisotropic MR is attributed to the mixing of spin up and down states of d-bands at the Fermi level due to SOI. This mixing is not isotropic because the magnetization direction provides a privilege axis for S-O perturbation which leads to the mechanism for anisotropic magnetoresistance.

S. D. thanks IIT Kanpur for financial help. This research work has been funded by IIT Kanpur, CSIR-India and DST-India. R. C. B thanks J C Bose National fellowship of the Department of Science and Technology. Authors thank Dr. H. Pandey for his help in fitting the low angle X-ray reflectivity data.

\end{document}